\documentclass[twocolumn,pre,showpacs]{revtex4}
\usepackage{graphicx}
\usepackage{amsmath}
\usepackage{amsfonts}
\usepackage{mathtools}

\begin{document}
\title{Unifying Three Perspectives on Information Processing in Stochastic Thermodynamics}

\author{A. C. Barato and U. Seifert}
\affiliation{ II. Institut f\"ur Theoretische Physik, Universit\"at Stuttgart, 70550 Stuttgart, Germany}

\parskip 1mm
\def\d{{\rm d}}
\def\Ps{{P_{\scriptscriptstyle \hspace{-0.3mm} s}}}
\def\MF{{\mbox{\tiny \rm \hspace{-0.3mm} MF}}}
\def\i{\text{\scriptsize $\cal{I}$}} 

\begin{abstract}
So far, feedback-driven systems have been discussed using 
(i) measurement and control, (ii) a tape interacting with 
a system or (iii) by identifying an implicit Maxwell demon 
in steady state transport. We derive the corresponding 
second laws from one master fluctuation theorem and discuss 
their relationship. In particular, we  show 
that both the entropy production involving mutual information 
between system and controller and the one involving a 
Shannon entropy difference of an information reservoir 
like a tape carry an extra term different from the usual 
current times affinity. We thus generalize stochastic thermodynamics to the presence of
an information reservoir.

\end{abstract}

\maketitle


A deep relation between information theory and statistical physics has been
apparent from the very conception of the former in Shannon's 
classical formulation \cite{shan48,jayn57}. One explicit manifestation is Bennett's insight on how
Landauer's result on the thermodynamic cost of erasing memory exorcises
Maxwell's demon \cite{benn82,maru09}. While thus the universal validity of the second law has
apparently been restored, exploring the specific relationship between information 
theory and thermodynamics particularly in small systems has become a very
active field not the least since ingenious experiments with single 
colloidal particles provide beautiful illustrations and test of these concepts \cite{toya10a,beru12}.
If the arguably bewildering plethora of recent theoretical work in this 
field \cite{touc00,touc04,cao04,cao09,saga10,ponm10,horo10,horo11,horo11a,gran11,abre11,abre11a,kund12,saga12,saga12b,baue12,
kish12,mand12,bara13,mand13,espo12b,stra13,horo13,gran13,kawa07,espo11,andr08,andr13} is tentatively classified into three main approaches
an important question on the uniqueness of the second law arises as follows. 

In the first and most prominent approach, the classical ideas of Maxwell and Szilard are 
implemented in an explicit feedback scheme where immediately after 
a measurement  some parameters of the device are 
altered depending on the outcome of the measurement \cite{touc00,touc04,cao04,cao09,saga10,ponm10,horo10,horo11,horo11a,gran11,abre11,abre11a,kund12,saga12,saga12b,baue12,
kish12}. The subsequent
evolution of the system thus depends on the state after the measurement and 
the new control parameter. For such a set-up, Sagawa and Ueda have derived an
integral fluctuation theorem (FT) \cite{saga10}. The corresponding inequality implies that the extracted work, which from the perspective of the first law is
compensated by a corresponding heat transfer from the bath, is less than the 
information acquired in the measurement \cite{cao09}. This inequality thus generalizes the
second law to such feedback-driven schemes. Since the thermodynamic
cost of neither the measurement nor of the erasure of the acquired information are
included, the analysis is necessarily somewhat incomplete from a thermodynamic point of view.

A second approach where the system is allowed to interact explicitly with 
an information storage device such as a tape consisting of a sequence 
of bits overcomes this deficiency.   
In such a scheme, an inequality has 
been derived which shows that the work extracted from a heat bath is
necessarily less than the information theoretic entropy difference between
outgoing and incoming tape \cite{mand12} (see also \cite{bara13,mand13}).    
How is this inequality related to the one derived in the first approach? 
Does it also follow from an underlying fluctuation theorem?

In the third approach, ``ordinary'' transport through a device 
like a quantum dot controlled by a gate is considered \cite{espo12b,stra13}. The
corresponding non-equilibrium steady state (NESS)  complies with the well-established 
rules of stochastic thermodynamics including a well-defined rate of
thermodynamic entropy production \cite{seif12}. A {\sl posteriori}, however, a term in the entropy production
is interpreted as an ``information current'', which is related to an idealized 
feedback procedure happening much faster than the time-scales for the transitions between states.    
Again, the question arises whether and how the genuine thermodynamic entropy production of a NESS 
relates to the second laws discussed within the first two approaches. A recent work in
this direction compares this genuine entropy production with the one arising from the 
first approach by considering two models with similar dynamics that can be either driven by an input
of chemical work or by feedback \cite{horo13}.     

In this letter, we will show that the second laws arising from these three 
approaches are in fact three different inequalities involving three
different quantities each bounding the maximal extractable work from such
devices. We will do so by first discussing the simplest paradigmatic device
based on a two level system from all three perspectives.  For a system with an
arbitrary number of states, we then derive one master FT which can be
specialized to yield the three second laws pertaining to the three perspectives discussed above.
Based on these insights, we can thus generalize stochastic thermodynamics
to include an information reservoir, like a
a tape that mediates transitions between a pair of states in a general NESS. 
Surprisingly, differing from the usual thermodynamic entropy production which can be written
as a sum of currents multiplied by affinities,
the contribution due to the information reservoir does not involve a current.


Let us set the stage with a paradigmatic two level system \cite{espo10a,kuma11a,cao09,horo13}. The upper level $u$ has energy $E>0$ and the lower level $d$ has energy $0$. The system is connected to a heat bath at temperature $T$
so that the transition rates fulfill the detailed balance relation $k_+/k_-= \exp(-E)$, 
where we set Boltzmann's constant multiplied by the temperature to $k_BT\equiv1$, $k_+$ is the transition rate from $d$ to $u$, and $k_-$ is the reversed one. The feedback is introduced in the following way.
After every period $t$ a measurement gives information to a controller whether the state of the system is $d$ or $u$. If the measurement is error free and if at the end of the time interval the system is at $u$, 
the energy of the upper level is lowered to $0$, leading to
the extraction of work $E$. Furthermore, the energy of the empty state is elevated to $E$ at no cost. This instantaneous change in the energy levels of the system
corresponds to an effective jump from state $u$ to state $d$, because after the energy levels are switched the labels are also switched with $d$ always representing the state with energy
$0$ and $u$ the state with energy $E$.

More generally, we assume a probability of a measurement error given by $\epsilon$, so that if the state at time $t$ is $x=u$ ($x=d$) 
the measurement yields $y=u$ ($y=d$) with probability $1-\epsilon$ and $y=d$ ($y=u$) with probability $\epsilon$. Moreover, for 
$y=u$ the energy levels are interchanged and for $y=d$ they remain fixed. Note that whenever an error occurs the
initial state in the next time interval is $u$. Hence, the system reaches a periodic steady state for which the probability of finishing the period at state $u$ is 
\begin{equation}
p_t= p+(\epsilon-p)\exp(-kt),
\end{equation}
where $k\equiv k_++k_-$ and $p\equiv k_+/k$. The mean extracted work per time interval $t$ is given by
\begin{equation}
W_t= E[p_t(1-\epsilon)-(1-p_t)\epsilon]= E[p_t-\epsilon].
\label{extw}
\end{equation}  
The probability of the measurement outcome $y=u$ is
\begin{equation}
q_t= p_t(1-\epsilon)+(1-p_t)\epsilon= p_t+\epsilon(1-2 p_t),
\label{defqt}
\end{equation}  
while the probability for $y=d$ is $1-q_t$. Therefore, the Shannon entropy of the controller is $H_y= H(q_t)\equiv -q_t\ln q_t-(1-q_t)\ln(1-q_t)$. 
This Shannon entropy conditioned on the state of the system $x$ becomes $H_{y|x}=H(\epsilon)$ \cite{footnote}. Using
the standard definition for the mutual information $I_t$ between the system $x$ and the controller $y$ \cite{cove06}, we obtain $I_t\equiv H_y- H_{y|x}=  H(q_t)- H(\epsilon)$.
The second law of thermodynamics for feedback controlled systems,  then implies \cite{cao09} 
\begin{equation}
I_t-W_t= H(q_t)-H(\epsilon)-W_t\ge0,
\label{2lawa}
\end{equation}
i.e., the extracted work is bounded by the mutual information due to measurements.  

\begin{figure}
\includegraphics[width=82mm]{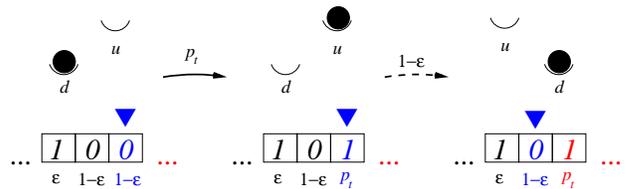}
\vspace{-2mm}
\caption{Two state model with feedback at fixed time intervals interpreted as a tape interacting with a heat bath and a work reservoir. The full arrow represents 
an interaction time interval $t$ and the dotted arrow an instantaneous effective transition due to the new incoming bit.
}
\label{fig1} 
\end{figure}

This very model allows a second interpretation which leads to another second law like inequality, see Fig. \ref{fig1}. 
We now consider a system connected to a thermal bath, mediating the interaction between 
a work reservoir and a tape (a sequence of bits), which  corresponds to  a simplified version of the original model proposed in \cite{mand12}. 
In this interpretation each bit from the tape interacts for a time $t$ with the system.  
During this time interval the bit state $0$ ($1$) is coupled to the system state $d$ ($u$), so that when the system jumps from
$d$ ($u$) to $u$ ($d$) the bit changes from $0$ ($1$) to $1$ ($0$). After interacting with the system for a time $t$, the bit moves forward, and a new bit comes to interact with the tape.
This new incoming bit generates effective transitions by determining the initial state of the system for the subsequent interaction time interval, where for an incoming $0$ the system will start at $d$ and for an incoming $1$ at $u$.
More precisely, if the system finishes in state $u$ ($d$) and the new incoming bit is $0$ ($1$) then the energy levels are interchanged, 
leading to an amount of energy $E$ extracted from (delivered to) the system. If the system finishes in state $d$ ($u$) and the new incoming bit is $0$ ($1$) then the energy levels remain fixed and no energy is
exchanged with the work reservoir. Furthermore, the probability of an incoming $1$ is $\epsilon$ and hence the Shannon entropy of the incoming tape is $H(\epsilon)$. On the other hand, the outgoing tape is a record
of the interaction with the system, with the probability of a $1$ being $p_t$ and the Shannon entropy $H(p_t)$. Importantly, in this second interpretation, we have an autonomous system with no explicit measurement and feedback,
the new incoming bit simply determines the initial state of the system for the coming interaction period.
  
Mandal and Jarzynski \cite{mand12} showed that a second law like inequality bounds the work (\ref{extw})  delivered to the reservoir 
by the Shannon entropy difference between the incoming and the outgoing tapes, i.e., 
\begin{equation}
H(p_t)-H(\epsilon)-W_t=D_{KL}(\epsilon||p)-D_{KL}(p_t||p)\ge0,
\label{2lawb}
\end{equation}
where $D_{KL}(x||y)\equiv x\ln(x/y)+(1-x)\ln((1-x)/(1-y))$ is the Kullback-Leibler distance. Hence, as our first result, we realize that this two level system allows for two different interpretations leading to two different 
second law like inequalities. Since $H(p_t)\le H(q_t)$ \cite{footnote}, the bound on the extracted work (\ref{2lawb}) is tighter than (\ref{2lawa}). Moreover, while the mutual information $I_t$ is always non-negative, $H(p_t)-H(\epsilon)$ 
can be negative. Hence, for $p_t<\epsilon<1/2$ the work delivered to the system $-W_t$ is used to erase information, with the reduction of the Shannon entropy of the tape being
bounded by $-W_t$, as given by (\ref{2lawb}) (see \cite{mand12,bara13}). Such information erasure cannot be addressed within the second law inequality (\ref{2lawa}) as $I_t\ge 0$.

Preparing for a third perspective on this model as a NESS, we assume that the feedback procedure does not take place at constant time intervals $t$ but that it is rather a Poisson process with rate $\gamma$.
Consequently, the previous expressions obtained for a fixed $t$ must be weighted with $\mathrm{e}^{-\gamma t}$. The average extracted work then becomes with (\ref{extw})
\begin{equation}
W_\tau\equiv \gamma \int_0^\infty dt \exp(-\gamma t) W_t= E(p_\tau -\epsilon)
\end{equation}
where $\tau\equiv k/(k+\gamma)$ and $p_\tau\equiv \tau p+(1-\tau) \epsilon$. Using the inequality
$\int_0^\infty dt \gamma\exp(-\gamma t) H(p_t)\le H(p_\tau)$ for the concave function  $H(x)$, the second law inequality (\ref{2lawb}) can be written in the form
\begin{equation}
\dot{s}_1\equiv \gamma(H(p_\tau)-H(\epsilon)-W_\tau)\ge0,
\label{2lawconta}
\end{equation}
where $\dot{s}_1$ represents a rate of entropy production. Analogously, the inequality involving the mutual information (\ref{2lawa}) becomes
\begin{equation}
\dot{s}_2\equiv \gamma(H(q_\tau)-H(\epsilon)-W_\tau)\ge0,
\label{2lawcontb}
\end{equation}
where $q_\tau= p_\tau+\epsilon(1-2 p_\tau)$.

The NESS description of this model then follows  by considering two states $d$ and $u$ with two links between them. One link is related to the thermal reservoir and the corresponding transitions rates are $k_+$ and $k_-$ as before. 
The other link is related to the effective transitions mediated by the tape with transition rates $\gamma\epsilon$ and $\gamma(1-\epsilon)$.  
The master equation for this model is analogous to the master equation for the previous model with feedback. More precisely, the stationary state probability distribution is $P_u= p_\tau$. 
The rate at which work is delivered to the mechanical reservoir is $\dot{w}= \gamma W_\tau$. The usual rate of thermodynamic entropy production specialized
to this NESS becomes \cite{seif12}  
\begin{align}
\dot{s} & = k[p_\tau(1-p)-(1-p_\tau)p]\ln\frac{1-p}{p}\nonumber\\
& +\gamma[p_\tau(1-\epsilon)-(1-p_\tau)\epsilon]\ln\frac{1-\epsilon}{\epsilon}\nonumber\\
& = \gamma[(p_\tau-\epsilon)\ln\frac{1-\epsilon}{\epsilon}-W_\tau]\ge 0.
\label{usual2law}
\end{align} 
For $\epsilon \to 0$ this thermodynamic entropy production
diverges in contrast to both (\ref{2lawcontb}), for which $\epsilon=0$ implies error free measurements, and (\ref{2lawconta})
for which $\epsilon=0$ means a fully ordered incoming tape. In the first case, the physical reason for this very different behavior 
 comes from the fact that (\ref{2lawcontb}) does not contain the thermodynamic cost of acquiring or erasing information \cite{gran11,saga12b,horo13}.
In the second case, a remarkable result is obtained if we compare (\ref{2lawconta}) with (\ref{usual2law}). Let us consider $\epsilon<p_\tau<1/2$, so that the
flow of work to the mechanical reservoir is positive. The minimal rate of work $\dot{w}_c$ that would have to be provided by the mechanical reservoir in order to restore the original tape 
(with a fraction $\epsilon$ of $1$'s) from the processed tape (with a fraction $p_\tau$ of $1's$) is obtained in the adiabatic limit $k\gg \gamma$ with $E=\ln\frac{1-\epsilon}{\epsilon}$. 
According to (\ref{2lawconta}), it is given by $\dot{w}_c=\gamma(p_\tau-\epsilon)\ln\frac{1-\epsilon}{\epsilon}\ge \gamma(H(p_\tau)-H(\epsilon))$.
Thus, if we apply (\ref{2lawconta}) twice, first for extracting work at the
expense of increasing the entropy in the tape and second for
restoring the original tape by applying mechanical work in the limit $k\gg \gamma$, we
find for the total entropy production the bound (\ref{usual2law}). Hence the NESS
description contains the full thermodynamic cost including the one for restoring the original tape. 
This observation shows that in a fully integrated
description, an error-free or perfect tape  scheme implies an
infinite thermodynamic cost somewhere else, as noted previously for a particular case study in \cite{stra13}.
 


\begin{figure}
\includegraphics[width=72mm]{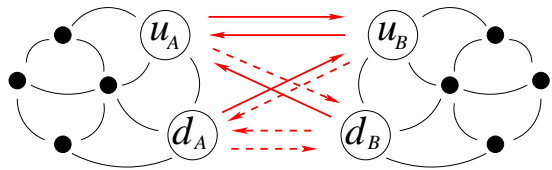}
\vspace{-2mm}
\caption{Representation of the formal duplication of the system. The full circles represent generic states $i$ different from $u$ and $d$; the curved lines
represent links between states with non-zero transition rates $W_{ij}$ and $W_{ji}$; the arrows represent the transition rates between the replicas, with the full arrows representing $\gamma \epsilon$ and 
the dotted arrows representing $\gamma (1-\epsilon)$.
}
\label{fig2} 
\end{figure}

Leaving the paradigmatic two state system we now derive a master fluctuation theorem for a general Markov process with transition rates from generic states $i$ to $j$ denoted by $W_{ij}$,
which will lead to the generalized version of the three entropy productions (\ref{2lawconta}), (\ref{2lawcontb}), and (\ref{usual2law}).
In stochastic thermodynamics the transitions rates are related to
reservoirs, with the ratio $W_{ij}/W_{ji}$ given by the local detailed balance condition \cite{seif12}. For simplicity, we consider the case where there is at most one link 
for each pair of states except for one pair. Denoting the two states of this special pair by $u$ and $d$, besides the ordinary transition rates $W_{ud}$ and $W_{du}$ (which can be zero), there are  
also rates $R_{ud}$ and $R_{du}$, which will become related to an information reservoir.

In order to derive the master FT it is convenient to formally duplicate the system, see Fig. \ref{fig2}. We represent the two copies of the system by the subscripts $A$ and $B$. The ``internal'' transition rates $W_{ij}$ are the same for both sides 
and they involve states with the same subscript, i.e., from $i_A$ to $j_A$ or from $i_B$ to $j_B$. The transition rates related to
the information reservoir, $R_{ud}=R_{dd}=\gamma(1-\epsilon)$ and $R_{du}=R_{uu}=\gamma\epsilon$, must 
involve states with different subscripts.  
This description is clearly symmetric with the stationary state probability distribution fulfilling  $P_{i_A}=P_{i_B}=P_i/2$, for all states $i$. Therefore,
the stationary properties of the duplicated system and of the original one are the same.

We denote a stochastic trajectory from time $0$ to $T$ with $N$ jumps visiting states $x_n$ (with $n=0,1\ldots,N$) by $X_T$. The master FT is derived by considering a reversed trajectory $\tilde{X}_T$ subjected to, in general, different transition
rates, denoted by an over-line, fulfilling the constraints $\overline{W}_{ij}= W_{ij}$ and $\overline{R}_{du}+\overline{R}_{dd}= \overline{R}_{ud}+\overline{R}_{uu}=\gamma$.
Considering the total internal current from $i$ to $j$
\begin{equation}
\mathcal{J}_{ij}[X_T]\equiv  \sum_{n=0}^{N}\sum_{C=A,B}(\delta_{x_n,i_C}\delta_{x_{n+1},j_C}-\delta_{x_n,j_C}\delta_{x_{n+1},i_C})
\end{equation}
and the counter of jumps from $i$ to $j$ between the two replicas  
\begin{equation}
\mathcal{K}_{ij}[X_T]\equiv \sum_{n=0}^{N}(\delta_{x_n,i_A}\delta_{x_{n+1},j_B}+\delta_{x_n,i_B}\delta_{x_{n+1},j_A}),
\end{equation}
we define the functional
\begin{equation}
\Omega[X_T]\equiv \sum_{i<j}\mathcal{J}_{ij}[X_T]\alpha_{ij}+\sum_{ij}{}^{'}\mathcal{K}_{ij}[X_T]\beta_{ij},   
\end{equation}
where $\alpha_{ij}\equiv \ln(W_{ij}/W_{ji})$ and $\beta_{ij}\equiv \ln(R_{ij}/\overline{R}_{ji})$. The first sum is over all pairs $ij$ with $i<j$ and the second constrained sum is over the 
states $i=u,d$ and $j=u,d$ ($\mathcal{K}_{ij}=0$  if $i\neq u,d$ or $j\neq u,d$).
As a main result we can show that $\Omega$ obeys the  integral FT $\langle \exp(-\Omega) \rangle=1$,  
which implies $\langle\Omega \rangle\ge 0$, with the brackets representing an average over all trajectories \cite{footnote}. 
From this inequality, the three second law inequalities can be derived from three different choices  of $\overline{R}$ as follows \cite{footnote}.

First, for $R=\bar{R}$, the well known standard rate of entropy production generalizing (\ref{usual2law}) follows as   
\begin{equation}
\dot{s}= \sum_{i<j}J_{ij}\alpha_{ij}+J'_{ud}\ln \frac{1-\epsilon}{\epsilon}\ge 0,
\label{usualrate} 
\end{equation}
where $J_{ij}\equiv P_iW_{ij}-P_jW_{ji}$ and $J'_{ud}\equiv \gamma(P_u+P_d)(p_\tau-\epsilon)$, with $p_\tau\equiv P_u/(P_u+P_d)$ . 

Second, choosing
\begin{equation} 
\overline{R}_{uu}=\overline{R}_{du}= \gamma p_\tau\textrm{ and }\overline{R}_{dd}=\overline{R}_{ud}= \gamma(1-p_\tau), 
\label{choice2}
\end{equation}
we obtain as the generalization of (\ref{2lawconta})
\begin{equation}
\dot{s}_1= \sum_{i<j}J_{ij}\alpha_{ij}+\gamma(P_u+P_d)[H(p_\tau)-H(\epsilon)]\ge 0,
\label{general} 
\end{equation}
This second law inequality generalizes the theory of stochastic 
thermodynamics to the presence of an information reservoir. In this entropy production, the term related to the transitions that are mediated by the tape is not the probability current $J'_{ud}$ multiplied by 
the affinity $\ln \frac{1-\epsilon}{\epsilon}$ that appears in the usual entropy rate (\ref{usualrate}). It is rather given by the rate at which the tape is processed, $\gamma(P_u+P_d)$, multiplied
by the Shannon entropy difference of the processed tape. Crucially, this quantity is not antisymmetric and therefore it is not subjected to the conservation laws of probability currents \cite{schn76,bara12a}.
This observation demonstrates that a formulation of the second law containing Shannon entropy differences related to information reservoirs is fundamentally different from the ordinary thermodynamic
entropy production. 


The physical meaning of the choice (\ref{choice2}) becomes clear if we consider the two state model again. The FT leading to
the inequality (\ref{2lawconta}) is obtained by considering a reversed trajectory where the probability of a $1$ in the incoming tape is $p_\tau$. If we go back to the initial model with feedback at fixed time intervals $t$,
our FT is obtained by applying feedback also to the reversed trajectory \cite{kund12}, however, the probability of an error for the reversed trajectory is chosen as $p_\tau$ rather than $\epsilon$. This is different
from the Sagawa-Ueda FT, where there is no feedback in the reversed trajectory \cite{horo10}.

Third and finally, the NESS version of the Sagawa-Ueda FT is obtained with transition rates $\overline{R}$ corresponding to a ``protocol'' in the reversed trajectory determined by 
the measurements along the forward trajectory \cite{horo10}. Therefore, with $\overline{R}_{ud}=\overline{R}_{du}= \gamma q_\tau$ and $\overline{R}_{uu}=\overline{R}_{dd}= \gamma(1- q_\tau)$, 
where $q_\tau= p_\tau+\epsilon(1-2 p_\tau)$, we obtain
\begin{equation}
\dot{s}_2= \sum_{i<j}J_{ij}\alpha_{ij}+\gamma(P_u+P_d)[H(q_\tau)-H(\epsilon)]\ge0, 
\end{equation}
which becomes (\ref{2lawcontb}) for the two state model. The particular term $\gamma(P_u+P_d)[H(q_\tau)-H(\epsilon)]$
in the entropy production $\dot{s}_2$ accounts for the mutual information between the system and the information reservoir. 

The three different entropy production obey the relations 
\begin{equation}
\dot{s}-\dot{s}_1= \gamma(P_u+P_d)D_{KL}(p_\tau||\epsilon)\ge 0
\label{ine1}
\end{equation}
and
\begin{equation}
\dot{s}_2-\dot{s}_1= \gamma(P_u+P_d)(H(q_\tau)-H(p_\tau))\ge 0,
\end{equation}
which show that $\dot{s}_1$ provides the tightest bound on $\sum_{i<j}J_{ij}\alpha_{ij}$. On the other hand, there is no general inequality between $\dot{s}$ and $\dot{s}_2$, as noted previously for the two state model in
the limit $k\gg\gamma$ in \cite{horo13}.

In conclusion, our unified perspective on three different
approaches to feedback-driven systems has revealed that
the corresponding expressions for entropy production are
genuinely different despite the fact that we could derive
all of them from one master FT. Significantly, both the
one containing the Shannon entropy difference of an
information reservoir like a tape of bits interacting with
the system and the one containing mutual information between a 
controller and the system cannot be written in the standard
form of a current times an affinity. This result points {\sl inter alia} to
a conceptual challenge for a future comprehensive linear response
theory of information processing. Apparently, an information
reservoir like a tape has features that are fundamentally
different from those of a heat or particle reservoir. Whether
allowing the tape to reverse its direction will suffice to
restore an ``ordinary'' thermodynamic behavior as found
in the case study \cite{bara13} remains to be seen. 
Finally, the second law inequality (\ref{general}) provides a general framework  
to study the entropic interaction between a tape and a thermodynamic system.
Two examples are the paradigmatic two state model where this entropic interaction generates
a flow of work to a mechanical reservoir (or lifts a falling mass \cite{mand12}) 
and a refrigerator powered by it \cite{mand13}.


Support by the ESF though the network EPSD  is gratefully acknowledged. We thank D. Hartich and D. Abreu for helpful discussions.




\eject

\section*{Supplemental material}

In Sec. \ref{sec1} we present details concerning equations (1-4) in the main text. Sec. \ref{sec2} contains the derivation of the master FT.
In Sec. \ref{sec3} we show explicitly how the three entropy productions for the two state model expressed in equations (7-9) in the main text follow
from the master FT.     

\section{Mutual information in the two state model with feedback}
\label{sec1}

The two state model can also be defined through the joint probability distribution
of the state of the system at the end of a period $x$ and the measurement $y$,
which becomes
\begin{equation}	
P(x,y)=\left\{
\begin{array}{ll} 
 (1-p_t)(1-\epsilon) & \quad \textrm{if $x=d$ and $y=d$}, \\
 (1-p_t)\epsilon & \quad \textrm{if $x=d$ and $y=u$}, \\
 p_t(1-\epsilon) & \quad \textrm{if $x=u$ and $y=u$}, \\
 p_t\epsilon & \quad \textrm{if $x=u$ and $y=d$}. 
\end{array}\right.\,
\label{Pxy}
\end{equation}
The marginals of the joint probability are $P(x)= \sum_{y}P(x,y)$ and $P(y)= \sum_{x}P(x,y)$. Explictly, they are given by
\begin{equation}	
P(x)=\left\{
\begin{array}{ll} 
 1-p_t & \quad \textrm{if $x=d$}, \\
 p_t & \quad \textrm{if $x=u$}, 
\end{array}\right.\,
\label{Px}
\end{equation}
and
\begin{equation}	
P(y)=\left\{
\begin{array}{ll} 
 1-q_t & \quad \textrm{if $y=d$}, \\
 q_t & \quad \textrm{if $y=u$}, 
\end{array}\right.\,
\label{Py}
\end{equation}
where $q_t= p_t+\epsilon(1-2p_t)$. The Shannon entropy associated with the measurements is defined as \cite{cove06}
\begin{equation}
H_y\equiv -\sum_y P_y\ln P_y= H(q_t). 
\end{equation}
Moreover, the conditional probability $P(y|x)= P(x,y)/P(x)$ follows from (\ref{Pxy}) and (\ref{Px}), i.e.,
\begin{equation}	
P(y|x)=\left\{
\begin{array}{ll} 
 \epsilon & \quad \textrm{if $x\neq y$}, \\
 1-\epsilon  & \quad \textrm{if $x=y$}. 
\end{array}\right.\,
\end{equation}
Hence, the conditional Shannon entropy reads \cite{cove06}
\begin{align}
H_{y|x} & \equiv  -\sum_{x,y} P(x,y)\ln P(y|x)\nonumber\\
& =-(1-p_t)(1-\epsilon)\ln (1-\epsilon)-(1-p_t)(\epsilon)\ln \epsilon\nonumber\\
&-p_t(1-\epsilon)\ln (1-\epsilon)-p_t\epsilon\ln \epsilon\nonumber\\
& = H(\epsilon). 
\end{align}
Finally, the mutual information between system and controller due to the measurements is \cite{cove06}
\begin{equation}
I_t\equiv \sum_{x,y}P(x,y) \ln \frac{P(x,y)}{P(x)P(y)}= H_y -H_{y|x}= H(q_t)-H(\epsilon). 
\end{equation}

In order to prove that the $I_t$ is larger than $H(p_t)-H(\epsilon)$ it is convenient to write
the probability $q_t$ in two forms,
\begin{align}
q_t &=  p_t+\epsilon(1-2p_t)\nonumber\\
& = 1-p_t+(1-\epsilon)(1-2(1-p_t)).
\end{align}
It is now easy to see that $p_t< 1/2$ implies $p_t\le q_t\le 1-p_t$ and $p_t> 1/2$ implies $1-p_t\le q_t\le p_t$. Since the Shannon entropy is symmetric and maximal  
at $1/2$, it follows that $H(q_t)\ge H(p_t)$, i.e., $I_t\ge H(p_t)-H(\epsilon)$.

\section{Derivation of the master FT and the three second law inequalities}
\label{sec2}

The probability of a stochastic trajectory running in time from $0$ to $T$ exhibiting $N$ jumps is given by 
\begin{equation}
\mathcal{P}[X_T]= P(x_0)\prod_{n=0}^{N-1}w_{n,n+1}\prod_{n=0}^N \exp(-\lambda_n\Delta t_n)
\end{equation}
where $P(x_0)$ is the initial probability distribution, $w_{n,n+1}$ is the transition rate from state $x_{n}$ to state $x_{n+1}$, $\Delta t_n$ is the waiting time in state $x_n$ and $\lambda_n$ the escape rate of state $x_n$. 
The probability of the reversed trajectory $\tilde{X}_T$ is denoted by      
\begin{equation}
\overline{\mathcal{P}}[\tilde{X}_T]= P'(x_N)\prod_{n=0}^{N-1}\overline{w}_{n+1,n}\prod_{n=0}^N \exp(-\overline{\lambda}(x_n)\Delta t_n),
\end{equation}
where $P'(x_N)$ is the initial probability distribution and the over-line indicates that for the reversed trajectory the transition rates are generally different. We choose over-line rates fulfilling the constraints 
$\overline{W}_{ij}= W_{ij}$ and
\begin{equation} 
\overline{R}_{du}+\overline{R}_{dd}= \overline{R}_{ud}+\overline{R}_{uu}=\gamma.
\label{constraint}
\end{equation}
Two functionals of the stochastic trajectory are important in the subsequent derivation, the
total internal current 
\begin{align}
\mathcal{J}_{ij}[X_T]\equiv & \sum_{n=0}^{N}(\delta_{x_n,i_A}\delta_{x_{n+1},j_A}-\delta_{x_n,j_A}\delta_{x_{n+1},i_A}\nonumber\\
& +\delta_{x_n,i_B}\delta_{x_{n+1},j_B}-\delta_{x_n,j_B}\delta_{x_{n+1},i_B})
\end{align}
and the counter of jumps between the replicas
\begin{equation}
\mathcal{K}_{ij}[X_T]\equiv \sum_{n=0}^{N}(\delta_{x_n,i_A}\delta_{x_{n+1},j_B}+\delta_{x_n,i_B}\delta_{x_{n+1},j_A}).
\end{equation}
The ratio of the probability of original and reversed trajectories then reads   
\begin{equation}
\frac{\mathcal{P}[X_T]}{\overline{\mathcal{P}}[\tilde{X}_T]}=\frac{P(x_0)}{P'(x_N)}\exp(\Omega[X_T]), 
\end{equation}
where
\begin{align}
\Omega[X_T]\equiv & \sum_{i<j}  \mathcal{J}_{ij}\alpha_{ij}+\mathcal{K}_{uu}\beta_{uu} \nonumber\\
  &+\mathcal{K}_{ud}\beta_{ud}+\mathcal{K}_{du}\beta_{du}+\mathcal{K}_{dd}\beta_{dd},
\end{align}
where $\alpha_{ij}\equiv \ln(W_{ij}/W_{ji})$ and $\beta_{ij}\equiv \ln(R_{ij}/\overline{R}_{ji})$.
Note that the waiting times cancel because the escape rates remain unaltered for the over-line transition rates with the constraint  (\ref{constraint}).
If we choose the initial probability distributions for the forward and reversed trajectories to be uniform, using standard methods \cite{seif12} we obtain the integral FT 
\begin{equation}
\langle \exp(-\Omega) \rangle=1,
\label{masterFT}
\end{equation} 
which implies $\langle\Omega \rangle\ge 0$, where the brackets indicates an average over all stochastic trajectories. 

In order to calculate the entropy rates we use 
\begin{equation}
J_{ij}\equiv\lim_{T\to\infty}\frac{1}{T}\langle {\mathcal{J}_{ij}}\rangle= P_i W_{ij}-P_j W_{ji},
\end{equation}
and
\begin{equation}
\lim_{T\to\infty}\frac{1}{T}\langle K_{ij}\rangle= \gamma P_i \epsilon_{j},
\end{equation}
where $\epsilon_u= (1-\epsilon_d)= \epsilon$ and $\epsilon_i=0$ for $i\neq u,d$. 
  
First, using the subscript $1$, we consider the case $\overline{R}_{du}=\overline{R}_{uu}= \gamma\overline{\epsilon}_1$ 
and $\overline{R}_{ud}=\overline{R}_{dd}= \gamma(1-\overline{\epsilon}_1)$,
obtaining the rate  
\begin{align}
 &\dot{\omega}_1 \equiv \lim_{T\to \infty} \frac{1}{T}\langle\Omega_1 \rangle\nonumber\\
& = \sum_{i<j}J_{ij}\alpha_{ij}+ \gamma(P_u+P_d)[H(p_\tau)-H(\epsilon)+ D_{KL}(p_\tau||\overline{\epsilon}_1)],
\label{ome1}
\end{align}
where $p_\tau= P_u/(P_u+P_d)$. This rate becomes $\dot{s}$ for $\overline{\epsilon}_1=\epsilon$ and $\dot{s}_1$ for $\overline{\epsilon}_1=p_\tau$, where the above rate achieves its minimal value. 

Second, by choosing $\overline{R}_{du}=\overline{R}_{ud}= \gamma\overline{\epsilon}_2$ and $\overline{R}_{dd}=\overline{R}_{uu}= \gamma(1-\overline{\epsilon}_2)$, we obtain, 
\begin{align}
& \dot{\omega}_2 \equiv \lim_{T\to \infty} \frac{1}{T}\langle\Omega_2 \rangle\nonumber\\
&= \sum_{i<j}J_{ij}\alpha_{ij}+  \gamma(P_u+P_d)[H(q_\tau)-H(\epsilon)+ D_{KL}(q_\tau||\overline{\epsilon}_2)],
\label{ome2}
\end{align}
where $q_\tau= p_\tau+\epsilon(1-2p_\tau)$.
The minimal value of this rate is achieved with the choice $\overline{\epsilon}_2= q_\tau$, which leads to the entropy rate $\dot{s}_2$.

\section{Explicit calculations for the two state model}
\label{sec3}

We calculate the three entropy productions for the two state model with the time periods drawn from a Poisson process at rate $\gamma$ analyzed in the first part of the paper
by specializing the general expressions (\ref{ome1}) and (\ref{ome2}).
In this case we have only the two states $u$ and $d$ with $W_{du}=kp$ and $W_{ud}= k(1-p)$. The stationary state solution of the master equation gives
\begin{equation}
P_u= \frac{W_{du}+R_{du}}{W_{du}+R_{du}+W_{ud}+R_{ud}}= \frac{k p+ \gamma \epsilon}{k+\gamma}.
\end{equation}
In this two state model, $P_u= p_\tau= 1-P_d$ and 
\begin{equation}
\sum_{i<j}J_{ij}\alpha_{ij}= J_{ud}\alpha_{ud}= -E(p_\tau-\epsilon)=-W_\tau,  
\end{equation}
where $E= \ln((1- p)/p)$. Hence, from equation (\ref{ome1}), with $\overline{\epsilon}_1=\epsilon$ we obtain 
\begin{equation}
\dot{s}=\gamma[(p_\tau-\epsilon)\ln\frac{1-\epsilon}{\epsilon}-W_\tau],
\end{equation}
which is equation 9 in the main part. Choosing $\overline{\epsilon}_1=p_\tau$ results in 
\begin{equation}
\dot{s}_1=\gamma(H(p_\tau)-H(\epsilon)-W_\tau),
\end{equation}
which is equation 7 in the main part. Furthermore, from (\ref{ome1}) with $\overline{\epsilon}_2= q_\tau$ we have
\begin{equation}
\dot{s}_2=\gamma(H(q_\tau)-H(\epsilon)-W_\tau),
\end{equation}
which is equation 8 in the main part.

\end{document}